\documentstyle[prl,preprint,aps]{revtex}
\begin{document}
\draft
\title{Quantum Chaos and Random Matrix Theory - Some New Results 
\\ {\small Presented at the workshop on  \\ ``Physics and Dynamics 
Between Chaos, Order, and Noise", Berlin, 1996}}
\author{U. Smilansky}
\address{{Department of  Physics of Complex Systems,} \\
{The Weizmann Institute of Science, Rehovot 76100, Israel}}
\date{\today }
\maketitle

\begin{abstract}
New insight into the correspondence between Quantum Chaos and Random
Matrix Theory is gained by developing a semiclassical theory for the
autocorrelation function of spectral determinants. We study in
particular the unitary operators which are the quantum versions of
area preserving maps. The relevant Random Matrix ensembles are the
Circular ensembles. The resulting semiclassical expressions depend on
the symmetry of the system with respect to time reversal, and on a
classical parameter $\mu = {\rm tr} U -1$ where $U$ is the classical
$1$-step evolution operator.  For system without time reversal
symmetry, we are able to reproduce the exact Random Matrix predictions
in the limit $\mu \to 0$.  For systems with time reversal symmetry we
can reproduce only some of the features of Random Matrix Theory. For
both classes we obtain the leading corrections in $\mu$. The
semiclassical theory for integrable systems is also developed,
resulting in expressions which reproduce the theory for the Poissonian
ensemble to leading order in the semiclassical limit.
 
\end{abstract}

\newpage

\section{Introduction}

 One of the most remarkable achievements of quantum chaology, was the
observation that spectra of systems which are chaotic in the classical
limit, obey universal statistics which follow the predictions of
Random Matrix Theory (RMT) \cite {BohigasLH}. In most cases, the above
correspondence was established by comparing a few statistical
functions, mostly two point densities, or functions derived from them
like the $\Delta_3$ or the $\Sigma^2$ statistics. The nearest neighbor
spacing distribution $P(s)$ is another example.  The study of n-points
$(n >2)$ spectral statistics which would test the quantum chaos - RMT
correspondence more closely, are much more difficult to implement in
practice.

In a previous paper \cite {KKU} we proposed a different measure, which
is based on the study of the statistical properties of the {\it
spectral determinant}, sometimes referred to as the {\it secular
function}, the {\it spectral $\zeta$ function} \/ or the {\it
characteristic polynomial}: It is the function which vanishes if and
only if its argument belongs to the energy spectrum. For matrices of
dimension $N$ the characteristic polynomial is
\begin {equation}
p_H(x) = \det (Ix-H) = \prod_{n=1}^N (x-\lambda_n(H))) = \sum_{l=0}^N
a_l(H)x^l .
\end{equation}
The statistical properties of this function, can be expressed in terms
of the statistics of either the eigenvalues $\lambda_n$ or the
coefficients $a_l$.  Since the two sets of variables are functionally
related, they are statistically equivalent.  In practice, however, one
cannot check the full spectral distribution, and therefore it is
advantageous to study statistical measures which are based on other
accessible quantities. The measure which was the subject of \cite
{KKU} was the autocorrelation function
\begin {equation}
C(\xi) = {\cal {N}} \left <\int_{-\infty}^{+\infty} w(x) p_H(x+\xi/2)
p_H(x-\xi/2) {\rm d} x\right >_H
\end{equation}
where $w(x)$ is a positive function with a finite support and unit
integral. $\cal{N}$ is a normalization constant so that
$C(0)=1$. Analytical expressions for $C(\xi)$ were derived in \cite
{haake} and \cite {KKU} for the standard RMT ensembles.  These results
can also be written down in terms of $\left <|a_l|^2 \right>_H $.
 
There are other advantage to the study of the statistics of the
spectral determinant, which are especially important for establishing
the connection between quantum chaos and RMT. The Gutzwiller trace
formula, which gives the semiclassical theory for the spectral
density, is not a proper function.  The derivation of spectral
statistics based on this theory, should therefore be augmented by
additional assumptions \cite {BerryA400} or truncation procedures
\cite {Bogolkeating}. In contrast, the semiclassical expression for the
spectral determinant involves a finite number of periodic orbits, and
therefore it converges on the real energy axis.  The semiclassical
spectral determinant preserves another important property, namely, it
is explicitly real for real energies (\cite {Bogomolny}, \cite
{DoronUS}, \cite {KeatingBerry}).  Thus, the semiclassical study of
the statistical properties of the spectral determinant can be based on
a relatively solid starting point. Last but not least, the
semiclassical spectral determinant shares many of its properties with
the Riemann Siegel expression for the Riemann $\zeta$ function on the
critical line. The autocorrelation function for the Riemann $\zeta$ is
the subject of a recent study of Cheung and Keating \cite
{Keatingzeta}.  The rigorous results obtained for the Riemann $\zeta$
case, provide some support to the physically reasonable, (yet
mathematically uncontrolled) approximations made in the semiclassical
theory to be discussed here.

 The behavior of $C(\xi)$ can be intuitively clarified, by considering
two extreme cases: An equally spaced (infinitely rigid) spectrum
produces a correlation function which is a strictly periodic function
of $\xi$,
\begin {equation}
C(\xi) = \cos \pi \xi
\end{equation}
Where $\xi$ is measured in units of the mean level spacing. On the
other hand, a Poissonian spectrum with $N$ spectral points, yields a
positive correlation which decays to zero on a scale which is
proportional to ${\sqrt N}$.  Thus, the lack of correlation between
the energy levels, induces a slowly decaying correlation function.
The canonical random ensembles all display level repulsion which
induce strong correlations.  It is expected, therefore, that the
``incipient crystalline character" \cite {Dyson} of the spectrum of
these canonical random matrix ensembles will manifest itself by
oscillatory, yet slowly decaying correlation functions. The more rigid
the spectrum, the more marked and persistent will be the oscillations
of $C(\xi)$.

In the present work we would like to concentrate on the spectra of
Unitary operators, and use the autocorrelation function of their
spectral determinant to study the correspondence between quantum chaos
and RMT.  The spectral statistics of Unitary operators appear
naturally in quantum chaology, when one studies, e.g., Floquet
operators corresponding to time periodic hamiltonians,
\cite {kicktop},\cite {DittrichUS} or scattering matrices \cite {USLH}.
 The quantization of classical area preserving mappings also involve
quantum unitary operators. \cite {baker} In cases where the classical
phase space is compact, the corresponding quantum Hilbert space is
finite, and then one deals with unitary matrices of finite
dimension. This is the case we shall study here. We shall provide a
new derivation of the results obtained in \cite {KKU}, and add some
new results.

 In the next chapter we shall define the correlation function for the
cases of interest here, and quote the RMT results which were obtained
recently by the Essen group \cite {haake}.  We shall then turn to the
semiclassical theory and assuming the system to be classically
integrable or chaotic, we shall establish the conditions under which
correspondence with Poissonian or RMT can be achieved.

\section {Some results from the Theory of Random Matrices}      

  The spectrum of a $N \times N$ unitary matrix $S$ consists of $N$
unimodular eigenvalues $e^{i\theta_l}, \ \ 1\leq l \leq N $.  It is
convenient to write the characteristic polynomial so that it is real
on the unit circle
\begin{equation}
Z_S(\omega)= e^{{i\over2}(N\omega - \Theta)} \det (I-e^{-i\omega}S)
\label {scatZ1}
\end{equation} 
Where $e^{i\Theta} = \det (-S) $. The characteristic polynomial can be
written down as
\begin{equation}
Z_S(\omega)= e^{{i\over2}(N\omega - \Theta)} \sum_{l=0}^N a_le^{-i\omega l}
\end{equation}
The unitarity of $S$ leads to the relations 
\begin{equation}
 e^{-i\Theta/2} a_l = e^{i\Theta/2} a_{\Lambda-l}^{*}
\label {scatZsymm} 
\end{equation}
The autocorrelation function reads now 
\begin{equation}
 C_{\beta} (\xi) = {\cal {N}}_{\beta} {1 \over 2\pi} \int_{0}^{2\pi}
\left <Z_S(\omega+\xi/2)Z_S(\omega-\xi/2) \right>_{\beta}{\rm d}\omega
 = {\sum_{l=0}^N <|a_l|^2>_{\beta} e^{i  \xi (l-{N\over 2})} \over
\sum_{n=0}^N <|a_n|^2>_{\beta}}
\label {corrcen}
\end{equation}
 where $< \cdot >_{\beta} $ stands for the average with respect to the
spectral measure of the circular ensembles of interest: the orthogonal
(COE, $\beta=1$), the unitary (CUE $\beta =2$) or the Poissonian
($\beta=0$) ensembles. The $<|a_l|^2>_{\beta}$ are the coefficients in
the Fourier expansion of the correlation functions. In the sequel we
shall develop the semiclassical theory for these coefficients.

 In the semiclassical derivation, we shall make use of the following
identities:
\begin{equation}
\det (I-x S) = \sum_{n=0}^N a_n x^n = \exp\left 
( -\sum_{k=1}^{\infty}{x^k\over k} s_k \right)
\end{equation} 
where $s_n= {\rm tr} S^n$. We define the generating function
\begin{equation}
G_{\beta}(x,y) = \left <\det(I-xS) \det (I-yS^{\dagger})\right
 >_{\beta} = \left < \exp\left ( -\sum_{k=1}^{\infty}{1\over k}(x^ k
 s_k+y^k s^{\ast}_k )
\right) \right >_{\beta}
\label{generating}
\end{equation} 
so that 
\begin {equation}
<a_n a^{\ast}_m>_{\beta} = \left .  {1\over n! m!} {\partial
 ^{n+m}\over \partial^n x\partial^m y} G_{\beta}(x,y) \right |_{x,y=0}
\label{acorr}
\end{equation}
 The most important message is that the building blocks of the
generating function are the traces $s_n= {\rm tr} S^n$.

 The ensemble averages $< |a_n|^2 >_{\beta} $ were calculated in \cite
 {haake} for all values of $\beta$:

\begin{equation}
<|a_n|^2>_{\beta =1} =1 +{n(N-n) \over N+1} \ \ ; \ \ <|a_n|^2>_{\beta
=2} = 1.
\label {haake}
\end {equation}
For the Poissonian ensemble one gets \cite {haake}
\begin{equation}
<|a_n|^2>_{\beta =0} = \left( \begin{array}{c}N\\ n \end{array}
\right)
\label{poisson}
\end{equation}
It is convenient to introduce the scaled correlation length $\eta= \xi
{2\pi\over N}$ and in the limit of large $N$ the sums in (\ref
{corrcen}) can be approximated as integrals, to give

\begin {equation}
C_{\beta = 1}(\eta) = {3\over 2} \left ( {\sin\pi \eta
\over \pi \eta} + {1\over \pi^2}
{\partial ^2\over \partial \eta^2}{\sin \pi \eta \over \pi \eta}\right) \
 \ \ ;  \ \ \
C_{\beta = 2}(\eta) = {\sin\pi \eta \over \pi \eta }
\label {corrcircle}
\end {equation}
The correlation function for the Poisson ensemble reads
\begin{equation}
C_{\beta = 0}(\eta) = \cos ^N {2 \pi\over N} \eta \approx \left (1-
{1\over 2}({2 \pi\over N} \eta)^2 \right)^N \approx
\exp \left (-{2 \pi^2 \eta ^2 \over N}\right ).
\label {corpoison}
\end{equation}
The Poissonian correlation function decays much slower than the
correlation functions of the other ensembles.

\section {The Semiclassical Theory}

 The quantum unitary operator $\hat S$ which we consider, is assumed
to be the quantum analogue of an area preserving map $\cal{M}$ acting
on a finite phase space domain of area $A$. (For the present
discussion we shall confine our attention to maps with the twist
property.  The semiclassical treatment can be extended to the general
case).  The phase space coordinates are denoted by $\gamma =
(q,p)$. The map can be defined in terms of a generating function which
is the action $ \Phi(q,q') $
\begin{equation}
p= - {\partial\Phi(q,q') \over \partial q} \ \ \ ; \ \ \ p'=
{\partial\Phi(q,q') \over \partial q'}.
\label{classmap} 
\end{equation}
The explicit mapping function $\gamma'= \cal {M} (\gamma) $ is
obtained by solving the implicit relations (\ref {classmap}).

The semiclassical expression for the matrix elements of $\hat S $ in
the $q$ representation is
\begin{equation}
<q|S|q'> =\left ({1 \over2\pi i}\right ) ^{1\over 2}
\left [{ \partial^2 \Phi(q,q') \over \partial q 
\partial q'}\right ] ^{1\over 2}
{\rm e} ^{ i \Phi(q,q') /\hbar}
\label {sclsmatrix}
\end{equation}
     
In the semiclassical limit, $\Lambda$, the dimension of the Hilbert
space where $\hat S$ acts, is the integer part of $ {A\over 2\pi
\hbar}$.

 We have seen above, that one can express the coefficients $a_l$ in
terms of the traces $s_n = {\rm tr}S^n$. The semiclassical
approximation for $s_n$ involves the periodic manifolds of the
classical map.  For hyperbolic maps \cite {tabor},
\begin{equation}
s_n = {\rm tr} S^n \approx \sum_{t\in P_n} {g_t n_t e^{ir
              (\Phi_t/\hbar-\nu_t {\pi\over 2})} \over |\det
              (I-M_t^r)|^{{1\over 2}} }
\label{trsemicl}
\end{equation}
$P_n$ is the set of all primitive periodic orbits of {\cal{M}}, with
 periods $n_t$ which are divisors of $n$, so that $ n=n_t r$. Orbits
 which are related by a discrete symmetry are counted once, and their
 multiplicity is denoted by $g_t$. $M_t$ is the monodromy matrix and
 $\nu _t$ is the Maslov index. The action along the periodic orbit is
\begin{equation}
 \Phi_t = \sum_{j=1}^{n_t} \Phi(q_j,q_{j+1}) \ \ \ ({\rm with} \ \
q_{n_t +1} = q_1 ).
\end {equation}

 For integrable maps, we use the phase space variables $(I,\phi)$
where $I$ is the classical invariant. The domain of the mapping is
$I\in [I_{min},I_{max}] ,
\phi \in [0,2\pi]$. In this representation, the action takes the form
$\Phi(\phi,\phi')=\Phi(\phi'-\phi)$. The explicit form of the map is
\begin{equation}
I'=I \ \ \ ; \ \ \ \Delta \phi = \phi'-\phi = F(I)
\end{equation}
Where $F(I)$ is the solution of the generating relation $I = {{\rm d}
\Phi(\Delta \phi)\over{\rm d} \Delta \phi}$.  The expression for ${\rm
tr} S^n$ is now
\begin{equation}
{\rm tr} S^n= s_n\approx \sum _{m=1}^n \left [ {2\pi \over n \hbar
F'(I_{n,m})}\right ]^{1/2} {\rm e}^ { i \left[ n\Phi(\Delta\phi=2\pi
{m\over n})/\hbar -( n+{1\over 2}) {\pi \over 2}
\right ] } 
\label{trinteg}
\end{equation}
Where the summation is carried over the periodic manifolds of period
$n$ and winding number $m$. They occur at values of $I$ which satisfy
$F(I_{n,m}) = 2\pi {m\over n}$.
  
 Before we can proceed any further, we must clarify an essential
point. In contrast with the RMT, the semiclassical theory deals with a
{\em single} system, which does not depend on any random parameter
which could represent an ensemble. However, averaging is mandatory in
order to get a meaningful theory since the quantities we calculate
fluctuate appreciably.  We generate the ``semiclassical ensemble" of
$s_n$, (with $1\le n \le \Lambda/2$) by considering the inverse Planck
constant as a parameter, and different realizations of the ensemble
are distinguished by the value of the parameter $\hbar^{-1}$. We
restrict $\hbar^{-1}$ to the interval $|\hbar ^{-1}-\hbar^{-1}_0| <
\Delta$ . The mean value $\hbar^{-1}_0$ is large enough to justify the
use of the semiclassical approximation.  That is, for typical orbits
$\hbar^{-1}_0|\Phi_t-\Phi_{t'}| >> 1$. The interval $\Delta$ is taken
to be small on the scale of $\hbar^{-1}_0$, but sufficiently large so
that $\Delta |\Phi_t-\Phi_{t'}| > 2\pi$. In this way, the phases
(mod$2\pi$) of the semiclassical expressions can be considered random.
The averaging over the "semiclassical ensemble" is effected by
\begin{equation}
< A > _{\hbar} = {1\over \Delta }\int_{\hbar^{-1}_0-\Delta/2}
^{\hbar^{-1}_0+\Delta/2} {\rm d} \hbar^{-1} A (\hbar^{-1}).
\end{equation}

With this definition of the ensemble average, we get for both
 classically integrable and chaotic maps,
\begin {equation}
 < s_n > _{\hbar} =0.
\end{equation}   
The variance for the classically chaotic case reads,
\begin{equation}
 < |s_n|^2 >_{\hbar} \approx \sum_{t\in P_n} { g_t^2 n_t^2 \over
|\det(I-M_t^r)|}
\label {diagonal}
\end{equation}
For integrable maps we get 
\begin{equation}
 < |s_n|^2 >_{\hbar} \approx {2\pi (I_{max}-I_{min}) \over 2\pi \hbar}
= \Lambda
\label {diagonalint}
\end{equation}
Thus, the $\hbar^{-1}$ averaging provides the well know diagonal
(random phase) approximation.  Due to the exponential proliferation of
periodic orbits in hyperbolic maps, action difference become smaller
as $n$ increase, which is the reason why the diagonal approximation is
not valid uniformly. We shall make use of the diagonal approximation
in the restricted range $n< \Lambda/2 \approx \hbar^{-1}_0$ where the
diagonal approximation is justified. For integrable maps, the variance
of $s_n$ is independent of $n$. The result ${1\over \Lambda}< |s_n|^2
>_{\hbar} \approx 1$ implies that the spectral two-point correlation
function for integrable systems is Poisson
\cite {BerryA400}.

 At this point we would like to make a crucial observation:

\noindent
$(\spadesuit) \ \ \ ${\em The semiclassical ensemble of $ \left\{ s_n
  \right\} $ is an ensemble of independent random Gaussian variables.}

We shall show that this is approximately valid as long as the values
of $n$ are kept bellow the dimension $\Lambda$ which is of order
$\hbar^{-1}_0$.  Consider the correlator $< (s_n)^k (s_m^{\ast})^l
>_{\hbar}$. If $n$ and $m$ are relatively prime, the actions $\Phi_t$
which contribute to $s_n$ and $s_m$ are sufficiently different. Thus,
all the terms in the product $(s_n)^k (s_m^{\ast})^l$ are oscillatory
and will yield a vanishing result upon averaging.  If $n$ and $m$ have
a common divisor, $j$, choose $k=m/j,l=n/j $, and all the amplitudes
which involve repetitions of the primitive orbits of length $j$ will
contribute non oscillatory terms to the correlator. However, for
hyperbolic maps, the number of periodic orbits which involve
repetitions is exponentially smaller than the total number of periodic
orbits, and the statistical independence of the variables $s_n$ and
$s_m$ is ensured. The corresponding approximation is more difficult to
justify for integrable maps because the proliferation of periodic
manifolds is only algebraic.
 
If we check all other correlators using the approximation that
repetitions can be neglected, we find that the $s_n $ are Gaussian
random variables. It should be noted that RMT implies that in the
limit of large dimensions, (which coincides with the semiclassical
limit) the traces are random Gaussian variables \cite {haake}. This
property, which is shared by the statistical and the semiclassical
ensembles, has far reaching consequences, and it constitutes a strong
link between RMT and quantum chaos.

Having established the statistical properties of the semiclassical
ensemble we can calculate the generating function (\ref {generating})
for this ensemble.
\begin{eqnarray}
G_{\hbar}(x,y)&=& \left <\det(I-xS) \det (I-yS^{\dagger})\right
>_{\hbar} \\ \nonumber &=& \left < \exp\left (
-\sum_{k=1}^{\infty}{1\over k}(x^ k s_k+y^k s^{\ast}_k )
\right) \right >_{\hbar} \\ \nonumber
&=& \exp \left( \sum_{k=1}^{\infty} (xy)^k { <|s_k|^2>_{\hbar}\over
k^2} \right)
\label{generhbar1}
\end{eqnarray} 
so that 
\begin {equation}
<a_n a^{\ast}_m>_{\hbar} = \delta_{n,m} \left .  {1\over n!} {\partial
^{n}\over \partial^n v} Z_{\hbar}(v) \right |_{v=0}
\label{acorrhbar}
\end{equation}
with
\begin{eqnarray}
Z_{\hbar}(v)&=& \exp \left( \sum_{k=1}^{\infty}
<|s_k|^2>_{\hbar}{v^k\over k^2} \right)
\label{genhbar}
\end{eqnarray}
The coefficients in the Taylor expansion $Z_{\hbar}(v) =
1+\sum_{k=1}^{\infty} A_l v^l$ can be obtained by taking successive
derivatives of the two expressions for $Z_{\hbar}(v)$
\begin{equation}
A_l = {1\over l}\sum_{k=1}^{l} A_{l-k}{<|s_k|^2>_{\hbar}\over k}
\label{newtonlike} 
\end{equation}
  
We would like to emphasize  the following important points 

\noindent
$\bullet $ One can use (\ref {newtonlike}) to calculate $ <|a_l|^2
>_{\hbar}= A_l$ in the range $1 \le l \le \Lambda/2 $ {\em
exclusively}. We have to impose this restriction because the
derivation which leads to (\ref {newtonlike}) is not exact. The
approximations destroy the delicate balance between the
$<|s_k|^2>_{\hbar}$ which is necessary to maintain the symmetry $
<|a_l|^2 >_{\hbar}= <|a_{\Lambda-l}|^2 >_{\hbar}$ which follows from
unitarity. Because of the same reason, $A_l\ne 0$ for $l>\Lambda$,
which contradicts the basic ingredient of the theory, namely, that we
deal with a characteristic polynomial of order $\Lambda$.

\noindent
$\bullet $ Because of the reasons listed above,
\begin{equation}
 C(\xi) \ne {\rm e}^{-i\xi/2\Lambda} Z_{\hbar}({\rm e}^{i\xi})
\label {CnotZ}
\end{equation}  
 
\noindent
$\bullet $ The input necessary to compute $<|a_l|^2 >_{\hbar}$
consists of all the $<|s_k|^2>_{\hbar}$ with $k\le l \le \Lambda/2
$. Thus, the semiclassical ensemble of $s_n$ can be confined to the
range $n\le \Lambda/2 $ where the arguments which support the
conjecture $(\spadesuit)$ can be most trusted.

 The semiclassical expressions for $<|s_k|^2>_{\hbar}$, depend on the
underlying classical dynamics (\ref
{diagonal}),(\ref{diagonalint}). We shall treat the various cases
separately.

\subsection {Classical chaotic dynamics}
    
 It is instructive to consider the Fredholm determinant for the {\it
classical} evolution (Frobenius Peron) operator
\begin {equation}
U(\gamma,\gamma') = \delta (\gamma'-{\cal{M}}(\gamma)),
\end{equation}
where $\gamma$ is a phase space point in the domain of ${\cal{M}}$. A
straight forward integration gives
\begin{equation}
u_n = {\rm tr}U^n = \sum_{t\in P_n} {n_t g_t \over |\det(I-M_t^r)|}.
\end{equation}
Comparing this expression with (\ref {diagonal}), we may write
\begin{equation}
 < |s_n|^2 >_{\hbar} \approx \sum_{t\in P_n} { g_t^2 n_t^2 \over
|\det(I-M_t^r)|} = <g_tn_t>_{cl(n)} u_n \ \ ,
\label {diagonalus}
\end{equation} 
 where for any phase space function $\rho(\gamma)$, $<\rho>_{cl(n)}$
is the approximate phase-space average of $\rho$ obtained by sampling
$\rho$ on the set of periodic points of period $n$ with the
appropriate weights.  As $n\to \infty$, $<\rho>_{cl(n)}$ approaches
the true phase space average of $\rho$, since $u_n \to 1$ in this
limit \cite {hannay}. Defining
\begin{equation}
g = \lim  {<g_tn_t>_{cl(n)} \over n}
\end{equation}
we get
\begin{equation}
< |s_n|^2 >_{\hbar} \approx g n u_n
\end{equation}

Thus, (\ref {genhbar}) can be written as
\begin{eqnarray}
 \label{genhbar1} Z_{\hbar}(v) & = &\exp \left( \sum_{k=1}^{\infty}
<|s_k|^2>_{\hbar}{v^k \over k^2} \right)
\nonumber \\ 
&\approx & \exp \left( \sum_{k=1}^{\infty} g u_k{v^k \over k} \right)
\\ \nonumber &=& \exp \left (-g {\rm tr} \log ({\cal {I}} - v U)
\right ) \\ \nonumber &=& \left ( \det ({\cal {I}} - v U) \right
)^{-g} \\ \nonumber &=& (\zeta_{Ruelle}(v))^g \nonumber
\end{eqnarray}
Where $\zeta_{Ruelle} (z)$ is the Ruelle $\zeta$ function for the
 classical mapping ${\cal {M}}$. It is defined in terms of the
 Fredholm determinant of the classical evolution (Frobenius Peron)
 operator by
\begin{equation}
\zeta_{Ruelle}(z) = \left(\det ({\cal{I}}-z U) \right)^{-1}.
\label {zeta} 
\end{equation}
The relation (\ref {genhbar1}) is one of the central results of the
present work, and we shall come back to it after obtaining explicit
expressions for $<|a_m|^2>_{\hbar}$.
 
Starting with the second line in (\ref {genhbar1}) we can easily
derive the recursion relations
\begin{equation} \label {finalit}
<|a_m|^2>_{\hbar} = {g\over m} \sum_{k=1}^m<|a_{m-k}|^2>_{\hbar} u_k
\end{equation}
which should be solved with the initial condition $<|a_0|^2>_{\hbar}
=1$.

 Let consider systems which are strongly mixing and for which all
transients die out on a short time scale. In such cases, we may
replace the $u_k$ by their asymptotic value $u_k=1$ for all $k$.  The
recursion relation (\ref {finalit}) reads now 
\begin{equation} \label
{recurcue} <|a_m|^2>_{\hbar} = {g\over m}
\sum_{k=1}^m<|a_{m-k}|^2>_{\hbar}.  
\end{equation} 
This is solved by
\begin{equation}
<|a_m|^2>_{\hbar}= \left( \begin{array}{c}m+g-1\\ g-1 \end{array}
\right).
\end{equation}
as can be checked by direct substitution.  This solution can be used
only for $m <\Lambda/2$. In the range $m > \Lambda/2$ it should be
augmented by the identity
\begin {equation} \label {symmmm}
 < |a_l|^2>_{\hbar}= < |a_{\Lambda-l}|^2>_{\hbar}
\end{equation}
which is a direct consequence of unitarity. These results can be
compared with the results of RMT (\ref {haake}).

Systems {\it without} time reversal symmetry (TRS) have $g=1$, and
$<|a_m|^2>_{\hbar}=1$.  Thus, the semiclassical result coincides with
the prediction of the theory for for the CUE, (see (\ref {haake})),
$<|a_m|^2>_{\beta =2}=1$ .

 In chaotic systems {\it with} TRS, $g=2$ and
\begin{eqnarray}
\label {sclgoe}
< |a_l|^2>_{\hbar}\ \ \approx \ \ \left\{ \begin {array}{l} 1 + l \ \
 \ \ \ \ \ \ \ \ {\rm for }\ \ 1<l \leq \Lambda /2 \\ 1+\Lambda -l \ \
 \ \ {\rm for}\ \ \Lambda>l \geq \Lambda/2 \end{array} \right. .
\end{eqnarray}
 
 This expression does not reproduce the RMT result for the COE case
(\cite {haake})
\begin{equation}
< |a_l|^2>_{COE} \ \ = \ \ 1+l{\Lambda \over \Lambda +1} -l^2{1 \over
\Lambda +1}
\end{equation}
 However, for large $\Lambda$, where the semiclassical approximation
is justified, the semiclassical result agrees with the exact
expression in a domain of $l$ values of size $\sqrt{\Lambda}$ in the
vicinity of the end points of the $l$ interval, $l=0$ and $l=\Lambda$.
The deterioration of the quality of the agreement between the
semiclassical and the RMT expressions when TRS is imposed is typical,
and it is an enigma in the field of quantum chaos \cite {BerryA400}.

 So far, we discussed systems for which all transients die out on a
fast time scale, which was imposed by setting $u_k = {\rm tr} U^k=1$
for all $k$. This is possible only when one eigenvalue of $U$ is $1$
and all the rest vanish. In generic systems, the spectrum is not
degenerate in this extreme way. Rather, beside the eigenvalue $1$
which corresponds to the conserved phase- space measure, the spectrum
is in the interval $[0,1]$, and it accumulates at $0$.  The rate of
decay of transients is determined by the magnitude of the eigenvalues
of $U$ which are less then $1$. To get the leading correction due to
the non vanishing eigenvalues of $U$, one can expand the recursion
relation (\ref {finalit}) to first order in $ \mu = u_1-1 $.  One
obtains in this way recursion relations for the correction to
$<|a_m|^2>_{hbar}$.  They are particularly simple for the cases with
$g=1,2$ and the corrected coefficients are
\begin {eqnarray}
<|a_l|^2>_{hbar}& =& 1+ \mu \ \ \ \ \ \ \ \ \ \ \ {\rm for } \ \ g=1
                \\ \nonumber & =& 1+ l+2 \mu l \ \ \ \ {\rm for } \ \
                g=2
\label{transient}
\end{eqnarray} 
The symmetry $<|a_l|^2>_{hbar}=<|a_{\Lambda-l}|^2>_{hbar}$ should be
implemented for $l>\Lambda/2$.  Recently, the Essen group studied
numerically the variances of the coefficients of the characteristic
polynomial for the quantum kicked top {\cite {haake}. They checked
systems with and without TRS, and their numerical results show
systematic deviations from the RMT predictions which are consistent
with the expressions (\ref {transient}). The numerical results for the
case without TRS is particularly convincing.

\subsection {Classical Integrable dynamics}

 For integrable maps we have (\ref {diagonalint})
\begin{equation}
 < |s_n|^2 >_{\hbar} \approx \Lambda
\label {diagonalint1}
\end{equation}
 
 The resulting recursion relations for the coefficients
$<|a_m|^2>_{\hbar}$ are
\begin{equation}
 <|a_m|^2>_{\hbar} = {\Lambda \over m} \sum _{k=1}^m
{<|a_{m-k}|^2>_{\hbar}\over k}
\end{equation}

 We were not able to find a close form for the solution of this
equation. However, to leading order $<|a_m|^2>_{\hbar} \approx
{\Lambda^m \over m!}$ which coincides with the leading term of the
result for the Poisson ensemble (\ref {poisson}). As can be checked by
direct evaluation, the deviation between the semiclassical and the
exact values occurs already in the expression for
$<|a_{m=2}|^2>_{\hbar}$.

\section {Conclusions}

 In this work we derived the semiclassical theory for the
autocorrelation functions of spectral determinants, by obtaining
explicit expressions for their Fourier coefficients.  The
semiclassical theory provided two basic ingredients - it supported the
conjecture $(\spadesuit)$ that the semiclassical ensemble of $s_n$
form a random Gaussian ensemble, and, it provided the variances
$<|s_n|^2>_{\hbar}$ for the various cases under study. The comparison
with the corresponding RMT results is not uniformly successful, and we
would like to correlate the degree of success to the accuracy by which
the semiclassical theory provides the two ingredients mentioned above.

	For {\em chaotic} systems {\em without} TRS, the assumption
$(\spadesuit)$ is well founded, and the semiclassical expression for
$<|s_n|^2>_{\hbar}$ coincides with $<|s_n|^2>_{\beta=2}$.  Hence the
semiclassical theory for the autocorrelation coefficients matches
exactly the RMT result.

 For {\em chaotic} systems {\em with} TRS, the assumption
$(\spadesuit)$ is well founded, but the semiclassical expression for
$<|s_n|^2>_{\hbar}$ agrees with $<|s_n|^2>_{\beta=1}$ only in the low
n domain. Hence the semiclassical and the RMT results agree to leading
order only.

 For {integrable} system the assumption $(\spadesuit)$ is not so well
founded, but the semiclassical expression for $<|s_n|^2>_{\hbar}$
agrees with $<|s_n|^2>_{\beta=0}$, resulting again in agreement only
to leading order between the semiclassical and the RMT results.

 The last point which should be mentioned is the natural occurrence of
the classical Ruelle $\zeta$ in the theory. This phenomenon was
recently observed by several groups which studied the correspondence
between quantum chaos and RMT using other statistical measures \cite
{AAA},\cite {Bogolkeating}. We would like to emphasize, whoever, that
it is not correct to use $\zeta_{Ruelle}$ directly in the quantum
theory, (see (\ref {CnotZ})).  Rather, the quantum character of the
problem must be introduced by truncating the coefficients at
$l=\Lambda/2$, and imposing unitarity via the relation
$<|a_l|^2>=<|a_{\Lambda-l}|^2>$. Remembering that $\Lambda$ plays the
r\^ole of the Heisenberg time in the present theory, we can interpret
the above statement as the analogue of the ``Riemann Siegel
look-alike" \cite {KeatingBerry} symmetry which is a basic ingredient
in the semiclassical theory of spectral determinants.

\section{Acknowledgments}
This work was supported by the Minerva Center for Nonlinear Physics
and the Israeli Science Foundation. I would like to thank M. Opper and
D. Miller for their suggestions which led to reformulate the older
version of the theory and to bring it to its present state.  I am
indebted to F. Haake for allowing me to use his numerical data and to
J. Keating, and Z. Rudnik for stimulating discussions.

\end{document}